# Superluminal transmission is possible from now on

Zbigniew Andrzej Nowacki

*Computer Engineering Department, Technical University of Lodz,*
*Al. Politechniki 11, 90−924 Lodz, Poland.*
`Fax:   +48 606 560 005`
E-mail: nova@kis.p.lodz.pl

It is known that superluminal transmission of information and energy contradicts Einstein's relativity. Here we announce an unusual TOE called 'nature theory' in which impossible things become possible. We present the scheme of an apparatus for sending signals over arbitrarily large distances with speeds arbitrarily exceeding the light speed in vacuum. Introducing the notions of effective speed and reliability of superluminal devices, we encourage experimenters to set and break world records in this new branch. At the same time we outline a mechanism (termed 'particle encapsulation') owing to which nature theory remains Lorentz invariant and so consistent with experiments. From among other numerous applications of nature theory we discuss briefly local antigravitation and new computing machines, called 'vacuum computers', applying 'cat principle'. They are of great interest because should enable humans to overcome the Gödel-Turing barrier.

**12 pages, 1 figure**

In the celebrated 1905 paper [1] Einstein changed our approach to time and laid down a new foundation for all modern physical science. One of the consequences of his theory of relativity is that no signal or interaction can be transmitted faster than $c$, the light speed in vacuum. But since the mid$-$1960s physicists have been carrying out a number of experiments [2$-$16] indirectly demonstrating superluminal activity of Nature. As any competitive relativity theory has been never created, and all physics cannot be refuted, psychological (to some extent) barriers prevented experimenters and referees from sending and considering useful superluminal signals. The situation was, nonetheless, disturbing, so various explanations have been proposed. The most popular maintains that there is a fundamental level (heavens?) at which, unlike the observational one (earth?), immediate signals can be sent (by God being at rest?). However, a connected theory of both the levels would have to use Galileo and Lorentz transformations simultaneously, which seems to be inconsistent.

**Einstein antinomy**

The proof of the superluminal transmission impossibility in special relativity is well-known, but here we shall need its probabilistic version. Suppose that we have an apparatus for sending signals with a speed $c+\epsilon$. It can be used, according to the relativity principle, by any inertial experimenter and in any direction. Similarly as every physical device, our apparatus need not be completely reliable. For simplicity, we describe it by a probability matrix $p_{ij}$, where $i$ equal to 1 or 0 denotes correspondingly sending or not a signal, while $j$ — receiving a signal or not in a time interval of specified duration $T$. (Note that $p_{i0}+p_{i1}=1$.) We assume that

$$p_{01} < p_{11},$$

i.e., our device works at least somewhat. Suppose further that an experimenter $E$ travels from $(\boldsymbol{x},s)$ to $(\boldsymbol{y},w)$. During the journey he performs an experiment whose result $X$ has a probability $p_1$ distinct from 0 and 1. Being at $(\boldsymbol{y},w)$, $E$ sends a superluminal signal $S$ to $(\boldsymbol{r},t)$ (rigorously speaking, to its neighborhood with the time size $T$) if and only if $X$ has been obtained. By virtue of Lorentz transformations we may assume that $E'$ (a human or computer) at $(\boldsymbol{r}',t')$ moves in such a way that

$$\mid \boldsymbol{x}' - \boldsymbol{r}' \mid \leq (s'-t')(c+\epsilon). \qquad (1)$$

Thus it is possible that, immediately after getting $S$, $E'$ sends a superluminal signal $S'$ to $(\boldsymbol{x}',s')$. Therefore, if $E$ receives $S'$ at $(\boldsymbol{x},s)$, then he knows that the probability of obtaining $X$ in the starting experiment is equal to

$$p_2 = \frac{p_1(p_{11}^2 + p_{10}p_{01})}{p_1(p_{11}^2 + p_{10}p_{01}) + p_0 p_{01}(p_{00} + p_{11})},$$

where $p_0 = 1 - p_1$. The first contradiction consists in that $p_2 > p_1$. Moreover, $E$ is clever. He substitutes $p_2$ for $p_1$ and gets $p_3$ greater again. From continuity it follows that $\{p_n\}$ converges to 1. This signifies that Einstein's dream comes true; all probabilities, quantum physics, Las Vegas, etc., vanish. Seriously speaking, we have proved that any superluminal transmission, even at very low probabilities, refutes special relativity. And this blights hopes that the uncertainty of quantum phenomena can save Einsteinian physics. We see that quantum mechanics is at odds with it, so another relativity theory is needed. This task was very difficult, but lately [17] it has been at last performed.

**Superluminal technology**

This is a historical moment, my reader. In Fig. 1. you may see the outline of the first official apparatus for sending superluminal signals, a human activity that will be, in my opinion, common in the third millennium. To build the device you need nine mirrors (denoted by $M$), three beam splitters ($BS$), two clocks ($C$), one laser ($L$), one photodetector ($D$), one computer or at least button ($B$), two walls or beam stops ($W$) and, first of all, two similar nonlinear crystals ($DC$) able to split an input photon into two output ones with less frequencies (the so-called 'parametric down conversion' [18]). The crystals should be optically pumped by a coherent pump wave emitted by $L$ and split at $BS_1$. This can cause the conversion of input photons at one or both crystals, each with the emission of a signal particle $s_j$ and an idler $i_j$.

Consider first the situation without the line from $C_1$ to $BS_2$. As photon paths reaching either $BS_2$ or $BS_3$ are indistinguishable, particles, no matter whether they have been converted, interfere. This means that the remaining dashed lines are not used either. In particular, similarly as in the usual interferometer, no photon (above the vacuum level) is registered by $D$. Note that in this phenomenon there is a second-order as well as fourth-order interference, but the latter, being useless at superluminal transmission, is not visible owing to the geometry of Fig. 1.

Now suppose that one of experimenters presses (after, maybe, long hesitation) $B$. Then $C_1$ records time and raises (or lowers, knocks over, etc.) $BS_2$, using, e.g., an electromagnet. This causes that idler photons cease to interfere, and they are registered by detectors having 100% efficiency, i.e., walls. But this enables one, in principle, to determine whether a signal photon arriving at $BS_3$ comes from $DC_1$ or $DC_2$. (Applying beam stops instead of walls can be more practical because the absorption should occur rather quickly [17].) Thus signal particles stop interfering as well, and about 50% of them can be recorded by $D$. (That the possibility of distinguishing idler photons destroys the interference of signal ones was verified experimentally in a similar configuration [19].) Therefore, a useful signal has been sent from $B$ to $C_2$. Indeed, the second experimenter knows that, e.g., she has been asked out to dinner.

FIG. 1. The scheme of an apparatus for superluminal transmission of signals.

It remains to estimate the speed of this signal consisting of all the three main types of particles: tardyons (here electrons denoted by bold dash lines), luxons (photons), and tachyons. We assume, for simplicity, that for each $j$ the sizes of $DC_j$ and the distances between $M_{j1}$, $M_{j2}$, and $M_{j3}$ as well as between $W$ and $BS_2$ are infinitesimal. Thus $DC_2$, $BS_2$, and $BS_3$ constitute a right triangle whose legs are denoted by, of course, $s$ and $i$. As letting idler photons pass destroys the interference of merely their partners, we get the maximal speed

$$v_{max} = \frac{c\sqrt{s^2 + i^2}}{s - i}.$$

Strictly speaking, $v_{max}$ is the speed of virtual tachyons [17] carrying a piece of information on raising $BS_2$. It is easy to see that $v_{max}$ exceeds $c$ whenever $0 < i < s$. For example, in the geometry of Fig. 1. it equals $c\sqrt{5}$. Increasing $i$ you may obtain greater and greater speeds (arbitrarily large, theoretically), but you should not reach or exceed $s$. Indeed, no particle, even virtual, can move with the infinitesimal speed or backwards in time [17]. More precisely speaking, if $i > s$, then information can be transmitted rather from $D$ to $BS_2$, while in the case of $s = i$ both the beam splitters work independently. Note, however, that the probability of the last setting vanishes.

To calculate the minimal speed, denote by $T_0$ the total time of electronic parts except $B$, and — by $T$ the entire time of raising $BS_2$. As the reaction time of $D$ depends on the difference of radiant fluxes reaching it, we may assume that the time is approximately equal to $kr^2/I$, where $r$ is the distance between $BS_1$ and $M_1$, and $I$ — the intensity of $L$. Thus we get

$$v_{min} = \frac{\sqrt{s^2 + i^2}}{\frac{s-i}{c} + \frac{kr^2}{I} + T + T_0}.$$

This implies that a necessary condition for the superluminal transmission is

$$r\sqrt{2} > cT_0.$$

Therefore, increasing $r$ and $I$ you may start. Note that $T$ is not so essential here because some signal photons can be affected even when $BS_2$ is raised only in part. To take advantage of this possibility you should use a statistical procedure.

Let a computer perform an experiment consisting of cycles following one another repeatedly, each of which involves two stages. In the first stage (termed the *action period*) the electromagnet begins to raise $BS_2$, raises it completely, and finally lowers to the initial position. In the second stage (termed the *standby period*) $BS_2$ is entirely lowered all the time. The duration $a$ of the action period is fixed for all cycles, while — $s$ of the standby one is increased gradually. Let $V$ be a positive number, and $w$ — a nonnegative one dependent, in general, on $V$. Denote by $Q$ be the quantity of alarms raised during $(t, t+a+s]$ excluding $(t+r/V, t+r/V+w]$, and by $Q_0$ — during $(t, t+r/V]$, where $t$ is the initial instant of the cycle. Suppose that the averages of those quantities in all cycles performed so far are denoted by the same symbols correspondingly. Then by the *reliability* of the apparatus with the transmission speed $V$ we mean

$$R(V) = 1 - \lim_{s \to \infty} \frac{Qr}{Q_0 V(a + s - w)}$$

provided $Q$ does not vanish. (Of course, in practice it suffices to increase $s$ merely to some ceiling.) The least upper bound of the set $\{V: \ 0 < R(V)\}$ is called the *effective speed* $v_{eff}$ of the apparatus. If we have $R(V) = 1$, then the device works at this speed perfectly. The reliability equal 0 for all speeds signifies that it operates in an entirely accidental way, whereas a negative $R(V)$ — that we should decrease $V$.

It is recommended that in a session either $w$ vanishes or $r/V + w$ for all examined $V$ is constant. The quantity obtained in this way is termed the *waiting period*; in our case it may be set to zero. However, the above statistical procedure can be equally well applied to assess other superluminal methods, and then $w$ will be able to be essential. We believe that experimenters, especially working so far at superluminal propagation but not only, will construct quickly the apparatus of Fig. 1. or present other devices. It is easy to verify that

$$\frac{p_{01}(V)}{p_{11}(V)} \leq 1 - R(V).$$

Therefore, to refute special relativity it suffices to transmit signals with $v_{eff}$ exceeding $c$, i.e., $R(V) > 0$ for some $V > c$. We think, similarly as many physicists dealing with superluminal propagation [15, 20], that this is certain. But do not worry and be happy; in our opinion we may live in a great palace instead of a hut.

**Nature theory**

The approach presented in [17] has been called *nature theory* simply because its only model is the whole Nature. In this theory the foregoing impossibility proof (similarly as its standard version) is false and so termed 'Einstein's antinomy'. This signifies that the transmission of information and energy over arbitrarily large distances with speeds arbitrarily exceeding $c$ becomes admissible from now on. At the same time nature theory respects Lorentz transformations between inertial observers. This presentation makes it sound as if you could eat your cake and have it further, but that is a mathematical fact. Precisely speaking, we have proved Lorentz transformations with weak assumptions not causing contradictions.

It is worth pointing out that nature physics is the third theory using Lorentz transformations. Let us recall that the first of them, Lorentz's himself, was based on the aether, and said how matter interacted with it. The next, Einstein's, was based on the physical (homogeneous and isotropic) space, and dealt with interactions between matter and space-time. The present approach [17] is based on the vacuum, and involves solely interactions between particles. This is possible owing to the full quantization of the nature theory vacuum consisting of virtual particles. They were known of quantum electrodynamics, but in nature theory virtual particles have been defined in a much more general and perfect way. This causes that [17] is not only an alternative relativistic study, but also an attempt at explaining quantum phenomena, too difficult — as he himself admitted — for Einstein. For instance, virtual particles enable us to carry out the second quantization of wave (the first one was done by Planck and Einstein). As a result, we obtain, e.g., a causal explanation of the double-slit experiment and a fully objective wavefunction interpretation implying that of Born. We are also in a position to give an answer to Wheeler's recent question: 'How come the quantum?' It is, in our opinion, the virtual particle of nature theory, the least object existing physically.

Virtual particles cannot be (for a reason following from their definition [17]) observed in a direct way. But Fig. 1. shows that they can be included in signals and then affect our real world. Therefore, we assume that every signal consists of particles. Now we are in a position to explain why Einstein's impossibility proof cannot be restored to nature theory. This brilliant scholar recognized that two moving observers did not experience time and distances in the same way. But he did not notice (unhappily, for otherwise quantum physics would not be such a problem for him [21]) that, aside from time dilation and length contraction, there was one more relativistic effect: particle encapsulation. It consists in that moving observers do not see exactly the same particles. This causes, in particular, that if (1) holds, then $E'$ does not see, i.e., cannot receive $S$. (Let us admit that it is easy to say: 'does not see', while in reality there are a lot of technical difficulties overcome only in [17].) Thus the action of the device does not depend on sending any signal, that is, $p_{01} = p_{11}$. Hence we infer that superluminal transmission does not lead to any contradictions. Furthermore, we have proved that if $O$ sends a signal, and the relative speed of $O$ and $O'$ is sufficiently small, then $O'$ will receive this signal. This ensures that superluminal transmission is physically feasible.

The apparatus presented in Fig. 1. is designed for sending information, but superluminal transmission of energy is possible as well, even though we have proved that by virtue of Lorentz transformations the speeds of real particles cannot exceed $c$. (This proof was lacking in Einsteinian theory.) To transmit the energy of real particles, they have to be transformed into a field energy. It is a property of virtual particles (constituting fields), so it can be sent with any speed. Finally, the energy is transformed into real particles.

Particle encapsulation should not be any great surprise. Indeed, it is slightly similar to to the so-called 'many-worlds interpretation of quantum mechanics' [22−24]. However, the encapsulation causes branching rather particles than worlds. (We do not maintain that if you go to the cinema tonight, then there is another universe in which you will stay home.) And their coalescence is possible whenever the danger that the experimenter will come to know own future disappears. In particular, there is no encapsulation in a region unless superluminal signals are received in it; i.e., nature theory contains special relativity.

It should be pointed out that our theory is local, albeit in the most general sense of this word. This means that the behavior of a particle depends solely on particles being in its neighborhood. Note that this definition does not exclude faster-than-$c$ interactions and signals. Their speeds in our approach are unbounded and finite, while according to Newton they could be infinite, and according to Einstein they had to be bounded. And somewhat surprisingly, nature theory is causal and indeterministic (as a younger sister of quantum physics) at the same time. We are in a position to explain exactly what forces real particles to act in an indetermined fashion. Nevertheless, the postulates of nature theory may be changed under the influence of new experimental facts. Therefore, it is possible that in (rather distant) future our theory will become determined. One thing is certain: Bell's inequalities [25] do not stand in our way.

**Antigravitation**

There are three main application areas of nature theory:

- Superluminal transmission of information and energy.
- Local and controllable antigravitation.
- Vacuum computers.

The realization of merely one of them can change our life. Nonetheless, if at least one of them has not been realized, I will be the first to say that nature theory is unsuccessful.

Local and controllable antigravitation caused by antimatter explains the symmetry violation in decays of some particles [26] and what we see through telescopes at the sky. Of course, it is at odds with Einstein's principle of equivalence, but is consistent with quantum physics. [17] demonstrates that we need not return to Newton's principle with its accidental equality of gravitational and inertial mass of particles belonging to matter. Instead, we have the fundamental postulate of nature theory called the controllability principle. Owing to it we are able to distinguish between gravitational and inertial forces by means of local experiments, and at the same time there is exactly one component of unified charge, termed mass, responsible for inertial, gravitational, and even quantum effects. In our approach Einstein's famous formula has to be replaced by

$$E = \mid m \mid c^2.$$

Two vertical lines are essential here because the mass $m$ can be negative, whereas the energy $E$ cannot (in nature theory). We predict that the antigravitational action of antimatter will be soon corroborated by experiments performed in CERN.

**Vacuum computers and cat principle**

The quantized vacuum of nature theory enables us to introduce the notion of a vacuum computer. A bit of the RAM memory of semiconductor-based machines corresponds to an excess or shortage of real electrons in a region. Analogously, a bit of VAM (Vacuum Access Memory) being used by vacuum computers is defined via a measure of virtual particles with a definite unified charge being in a region. There are many sorts of vacuum computers; their logic gates can consist of real or virtual particles. We have proved [17] that some of those machines will be able to compute non-recursive functions [27].

Quark-vacuum processors, whose logic gates are made of virtual quarks, will be perhaps the most excellent machines constructed by humans at any time (whenever nature theory turns out to be true). Here the vacuum allows one to omit the Planck-Einstein law (stating [17] that there are at most finitely many real particles crossing a bounded four-dimensional region). This law implies, in fact, the Church-Turing thesis [27], but it is not the sole obstacle to computing non-recursive functions. Note that even if an ordinary computer worked with an exponentially increased speed, it could not perform any infinite calculation during a finite time for a simple reason: its hardware would then have to remain in an indetermined (not being even random) state, which is physically impossible. That is why quark-vacuum processors will apply a mechanism called 'cat principle'. It says that a bit can be written and a logic gate can be used at most once. Owing to it some infinite calculations will be able to be performed. Indeed, they will be reduced to setting physical fields in perfectly determined states.

A few years ago the promising development of artificial intelligence was disturbed by publishing [28]. As people's dream of constructing machines really thinking is so winsome, Penrose's book has been repeatedly being criticized, misrepresented, or simply left unsaid. But facts are implacable; Penrose (refining Lucas' argumentation [29]) has shown that no Turing-equivalent machine, even equipped with a random mechanism, can exactly simulate the human mind. This involves not only electronic or molecular [30] computers but quantum ones [31, 32] as well. Penrose's proof ought not to be disregarded because it is based, in fact, upon Gödel's incompleteness theorem [33]. A solution of this problem is given only in nature theory; we have proved that Penrose's reasoning does not embrace quark-vacuum processors. In reality, these beasts will be much better at thinking than humans. Thus we suggest realizing Hilbert's famous program by means of physics (i.e., the consistency of mathematics would follow from that of Nature). Another potential application is establishing the notion of truth in that portion of mathematics which deals with integers. And — last but not least — no planetoid will threaten us.

One may ask about the chance that those machines will be built. Obviously, vacuum computers are related to quantum ones; in both the cases one tries to use something existing outside our real world. However, computer science has been an algorithmic field so far, while quantum physics commissions Nature to perform difficult tasks without defining how she is to do them. Quantum physicists say apologetically that nobody knows how Nature does it. In many cases this is now false. Nature theory is an algorithmic science; nature theorists know or have to define what Nature does, how and why. So far knowledge — and not its lack — has been rewarded, and we hope that this tendency will be continued.

We encourage you to read an introduction to nature theory contained in [17], especially if you are a physicist, computer scientist, or mathematician, but not only. You will find there, among other things, the description of an unusual scientific conference that will be held about 3000. Invitations to it will be very difficult of attainment, but everyone has still got a chance.

## ACKNOWLEDGMENTS

I thank Prof. D. Sankowski for creating excellent conditions to writing this article.